\documentclass[pdflatex,sn-mathphys-num]{sn-jnl}% Math and Physical Sciences Numbered Reference Style
\usepackage{graphicx}%
\usepackage{graphicx}
\usepackage{caption}
\usepackage{subcaption}
\usepackage{multirow}%
\usepackage{amsmath,amssymb,amsfonts}%
\usepackage{amsthm}%
\usepackage{mathrsfs}%
\usepackage[title]{appendix}%
\usepackage{xcolor}%
\usepackage{textcomp}%
\usepackage{manyfoot}%
\usepackage{booktabs}%
\usepackage{algorithm}%
\usepackage{algorithmicx}%
\usepackage{algpseudocode}%
\usepackage{listings}%
\theoremstyle{thmstyleone}%
\theoremstyle{thmstyletwo}%

\theoremstyle{thmstylethree}%

\begin{document}

\title[Article Title]{Understanding Big Bang Nucleosynthesis:
the physics and the history}


\author[1,2]{\fnm{Michael S.} \sur{Turner}}\email{mturner@uchicago.edu}

%%\author[2,3]{\fnm{Second} \sur{Author}}\email{iiauthor@gmail.com}
%%\equalcont{These authors contributed equally to this work.}

%%\author[1,2]{\fnm{Third} \sur{Author}}\email{iiiauthor@gmail.com}
%%\equalcont{These authors contributed equally to this work.}

\affil[1]{\orgdiv{Kavli Institute for Cosmological Physics}, \orgname{University of Chicago}, \orgaddress{\street{5640 S. Ellis Ave}, \city{Chicago}, \postcode{60637-1433}, \state{IL}, \country{USA}}}

\affil[2]{\orgdiv{Department of Physics and Astronomy}, \orgname{University of California, Los Angeles}, \orgaddress{\street{Portola Plaza}, \city{Los Angeles}, \postcode{90095-1547}, \state{CA}, \country{USA}}}

%%\affil[3]{\orgdiv{Department}, \orgname{Organization}, \orgaddress{\street{Street}, \city{City}, \postcode{610101}, \state{State}, \country{Country}}}


\abstract{Big-bang nucleosynthesis (BBN) is a pillar of modern cosmology.  It was the first major test of the hot big bang model, and today remains its earliest and most precise test.  It all began with the trailblazing 1948 paper of Alpher, Bethe and Gamow, who proposed non-equilibrium nuclear processes in the early Universe and an early radiation-dominated phase to explain the abundances of all the chemical elements in the periodic table.  While their neutron-capture model was deeply flawed, it initiated the path to the modern theory of BBN and the discovery of the cosmic microwave background (CMB). In this paper I clarify the basic physics of BBN, adding some new insights, and describe how the modern theory developed.  I finish with a discussion of the pre-discovery predictions of the temperature of CMB, including the missteps and the missed opportunities.  The legacy of big bang nucleosynthesis is enormous, and it is yet another illustration of how foggy it can be at the frontiers of discovery.}


\maketitle

\section{Introduction}\label{sec1}
The landmark 1948 paper by Alpher, Bethe and Gamow \cite{abc, MSTRF} that started it all was motivated by the failure of equilibrium processes in stars to be able to account for the abundances of the chemical elements in the periodic table.\footnote{For a through review of this point, see Ref.~\cite{AlpherRMP}.} Their paper -- hereafter $\alpha\beta\gamma$ -- broke new ground with its addition of photons to the cosmic mix and attention to particle interactions (here, nuclear reactions) during the earliest moments.\footnote{It also broke new ground with the addition of Hans Bethe as an unknowing author, to create the $\alpha\beta\gamma$ moniker.}  Both are central features early Universe cosmology today.

Gamow's theory, as it was often referred to, attracted a lot of attention, in no small part because of his stature and big personality.\footnote{Further, the germ of the idea traces even further back to a sole-authored paper he wrote in 1946 \cite{Gamow46} and talks he gave.}  Of the two new features, radiation (photons) and nuclear reactions, the former opened the door to the study of the early, radiation-dominated phase of the Universe, while the latter eventually led to the modern theory of BBN. 

The $\alpha\beta\gamma$ paper was sparse in detail, and it is not clear that the effect of radiation upon the expansion of the Universe had been properly taken into account.  The 1948 {\it Nature} paper by Gamow seven months later \cite{gamow48} had more detail, and made very clear that Gamow was properly considering a radiation-dominated Universe.  It also included numerical results from Alpher's thesis \cite{alpher48} (yet to be published).  The {\it Nature} paper by Alpher and Herman that followed two weeks later \cite{alpherherman48} corrected some numerical errors and made their 5\,K temperature prediction for the relic radiation.  These 1948 papers \cite{abc,gamow48,alpher48,alpherherman48} collectively define what I will refer to throughout as Gamow's neutron-capture theory.

\subsection{Radiation phase}
Because the energy density of matter and radiation evolve differently with the cosmic scale factor $a(t)$, radiation evolves as $a^{-4}$ while matter evolves as $a^{-3}$,  a hot, radiation-dominated beginning is a  consequence of even a tiny photon component in the Universe. Today, photons (and relativistic neutrinos) contribute about 0.01\% of the total energy density, yet earlier than 60,000\,yrs photons and other relativistic particles dominated the energy content of the Universe.  

Further, early on, through rapid electromagnetic interactions, photons thermalize with matter, resulting in a blackbody distribution characterized by a temperature that evolves as $a^{-1}$.  As the Universe cools and becomes matter-dominated, the embers of the hot beginning remain with us today in the form of the cosmic microwave background (CMB), its blackbody spectrum undistorted by the expansion.  

The details of the early radiation dominated phase were all correct.  Radiation energy density evolving as $a^{-4}$, the dominance of radiation over matter early on, and scale factor evolving as $t^{1/2}$ during the early radiation-dominated phase.\footnote{Correctly taking account of radiation and realizing that there is a radiation dominated phase is a non-trivial matter, illustrated by a related discovery that made the Hydrogen bomb possible.  Ulam's contribution to the Teller-Ulam bomb was the realization that at high temperatures, radiation pressure can dominate that of matter.}  The only missing piece was the thermal bath of other relativistic particles including $e^\pm$ pairs and neutrinos and antineutrinos.

\subsection{Nuclear physics}

The basic nuclear physics underlying their paper -- {non-equilbrium} neutron capture -- was wrong and it took almost 20 years to straighten out, culminating in the seminal 1967 paper by Wagoner, Fowler and Hoyle \cite{WFH}.  Moreover, when all the dust settled, there is no significant nucleosynthesis beyond $^4$He:  only lightest elements have their origin in the big bang, the rest have to produced in stars and elsewhere later on.  Nonetheless, the prediction of a large primordial abundance of $^4$He was a crucial early success of the hot big bang model.

The emphasis on neutrons was well placed for two reasons.  First, with only protons, the first step in nucleosynthesis would be $p + p \rightarrow d + e^+ + \nu_e$.  This process, which proceeds via the weak interaction, is far too slow to be important in the early Universe.  In the sun, the timescale for this process is billions of years, which is why the sun burns Hydrogen for such a long time. No neutrons, no nucleosynthesis in the early Universe.

The second reason is that the easiest way to build up the heavier elements is by adding neutrons, not protons, because of the Coulomb repulsion of a proton and a nucleus.

By starting with 100\% neutron matter, $\alpha\beta\gamma$ maxed out on neutrons, letting the protons arise from neutron decays.  As it turns out, in the thermal bath that existed in the Universe seconds after the big bang, a pure neutron state is not possible; rather, the balance between protons and neutrons is regulated by the weak interactions and fixed by the temperature.

\subsection{Ahead}

In the next section, with the benefit of more than 70 years hindsight, 
I elucidate the basic physics of BBN, with some new insights.  Then I go on to discuss my reconstruction of the physics history, having read the key papers with an expert's eye, paying close attention to the thread that led to the modern theory.  

The penultimate section is devoted to clearing up some persistent misunderstandings about BBN  
and a discussion of how well the temperature of the CMB could have been predicted, knowing the correct BBN physics.  
The missed opportunity of predicting the CMB temperature rivals Einstein's missed opportunity for predicting the expansion of the Universe.

I write this paper as a practicing scientist who has written close to sixty papers on big bang nucleosynthesis, and not as a science historian.  Of course I do so at the risk of eliciting the ire of both communities, too much history for the scientist and too much science for the historian.  I am willing to take the risk because  I find the history of {the physics} underpinning this pillar of modern cosmology fascinating, and because I find a clear description of the physics and its history lacking in the current literature. 

\section{Physics of BBN}

\subsection{Big picture}
Today, BBN is a mature subject, with computer codes that take into account not only all the relevant nuclear and particle reactions but also many small corrections, producing results with better than 1\% precision \cite{precisionBBN}.  My purpose here is to elucidate the basic physics that underpins it, not the details. 

Like much of the history of the early Universe, the natural starting point for the discussion is thermal equilibrium.  This is because the high-densities and  energies in the early Universe result in particle interaction rates $\Gamma$ that almost always occur more rapidly than the rate at which the temperature is changing, $T \propto 1/a$, where $a(t)$ is the cosmic scale factor.
During the early ($t<60,000\,$yrs) history of the Universe, photons and other relativistic particles in the thermal are by far the dominant component of the total energy density, and the age and temperature are related by, $kT/$MeV\,$\simeq (t/{\rm sec})^{-1/2}$, which follows from the Friedmann equation, see the \S\ref{secA1} for more details.

Of course, if thermal equilibrium were the entire story, the Universe today would be very boring, comprised of the iron-group elements at at temperature of 3\,K with no nuclear free energy to power stars and life here on Earth!  The deviations from thermal equilibrium, even though relatively rare, are crucial.

 The top-level story of BBN is simple:  Before the epoch of nucleosynthesis ($kT \gg 1\,$MeV and $t \ll 1\,$sec), the energy density of the Universe was dominated by that of a thermal bath of particles (photons, $e^\pm$ pairs and neutrinos and antineutrinos), with rapid particle interactions maintaining thermal equilibrium.  Free nucleons were thermodynamically favored over nuclei and weak interactions keep the ratio of neutrons to protons at its thermal equilibrium value, $n/p = \exp [-\Delta mc^2/kT ]$, close to unity ($\Delta mc^2 = 1.293\,$MeV).  

After BBN ($kT\ll 0.01\,$MeV and $t\gg 10^4\,$sec), nuclei are thermodynamically favored, any free neutrons have long-since decayed (neutron mean lifetime $\tau_n \simeq 880\,$sec), and most importantly, nuclear reactions have ceased (``frozen out'') owing to Coulomb barriers between the charged nuclei present.  Nuclear physics does not become relevant again until the first stars form a billion years or so later.

\begin{figure}[h]
\center\includegraphics[width = \textwidth]{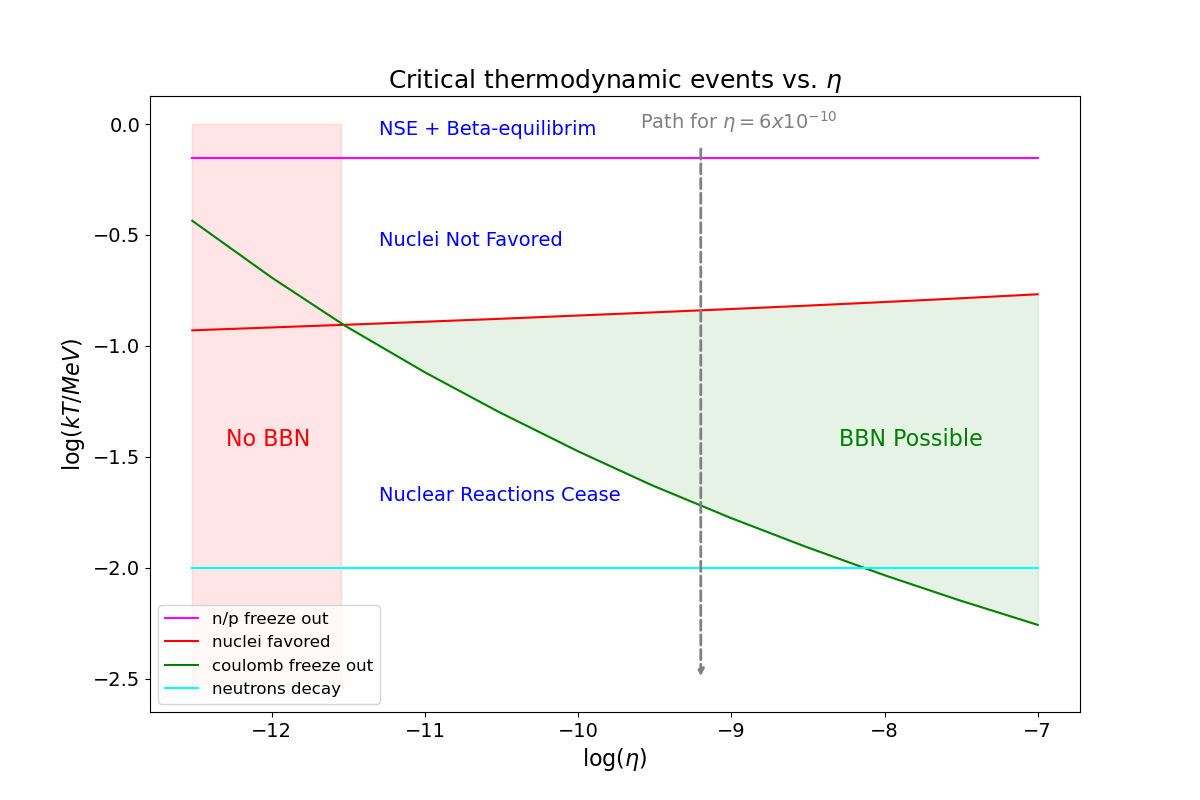}
\caption{Key thermodynamic events during BBN as a function of the baryon-to-photon ratio $\eta$:  $T_F$, the freeze in of the n/p ratio; $T_{\rm nuclei}$, the temperature below which nuclei are thermodynamically favored over free nucleons; $T_{\rm Coulomb}$, the temperature below which charged-particle nuclear reactions cease occurring; and $T_n$, the temperature at which the age of the Universe is the lifetime of a free neutron.  Any significant nucleosynthesis beyond deuterium requires $T_{\rm nuclei} \ge T_{\rm Coulomb}$, or $\eta \ge 10^{-11}$.  The gray vertical line shows the timeline of successful BBN:  freeze in of the n/p ratio at $T = T_F$;  a period of waiting until nuclei are favored $T_F > T > T_{\rm nuclei}$; nucleosynthesis $T_{\rm nuclei} > T > T_{\rm Coulomb}$; frozen-out nuclear reactions $T_{\rm Coulomb} > T$; and any free neutrons remaining decay $T_n > T$.}
\label{keyrates}     
\end{figure}

What happened in between involves two key departures from thermal equilibrium that determine the outcome of BBN. And this outcome only depends upon the baryon-to-photon ratio $\eta$, whose value remains constant as the Universe evolved and today is known to be $6.07 \pm 0.04 \times 10^{-10}$.\footnote{The baryon-to-photon ratio remains constant so long as there is no entropy creation or transfer because the number density of baryons decreases as $a^{-3}$ as does the number density of photons, which is proportional to $T^3$ and $T \propto a^{-1}$.  Around a temperature of $0.5\,$MeV, the entropy in $e^\pm$ pairs is transferred to the photons, increasing the photon number density by a factor of 11/4.  While the outcome of nucleosynthesis depends upon the temperature $T$ and the nucleon density $n_B$; because of the constancy of $\eta$, the nucleon number density is simply $n_N \propto \eta T^3$; see Refs.~\cite{Wagoner, KolbTurner}. The baryon-to-entropy ratio is exactly constant in the absence of significant entropy production, and today it is simply related to $\eta$:  $n_B/(s/k) = \eta /7.04$ \cite{KolbTurner}.}  

Strictly speaking, the results of BBN only depend upon $\eta$ if all the other input physics is fixed; this includes the number of neutrino species, any other low-mass particle species present, and all the relevant nuclear data (cross sections, lifetimes and so on).  While not all of the nuclear data is precisely known and there is the possibility of new physics beyond the Standard Model of particle physics. The Standard Model of particle physics is a good starting point and fixes everything beyond $\eta$.

\subsection{Baryon-to-photon ratio}
It is often said that the Hubble constant is the most important number in cosmology, setting the age and size of the Universe and defining the large-scale extent of the Universe today.  It can be argued that the baryon-to-photon ratio $\eta$ is the most important number for the physics of the hot big bang model.  Here is the case. 
 
 The inverse of $\eta$, or 2 billion photons per baryon, measures the radiative entropy.  This corresponds to an extraordinarily high entropy; for comparison, the value at the center of the sun is only around $10^{-3}$ photons per baryon and at the center of a newly-formed neutron star about 1 photon per baryon.   
 The fact that the entropy is so enormous means that some of the key physics of BBN seems counterintuitive.
 
The constancy of the baryon-to-photon ratio $\eta$ connects the temperature of the Universe and the baryon mass density at any epoch
\begin{eqnarray}
   \rho_B & = & \eta m n_\gamma = {2\zeta (3) (kT)^3 \over \pi^2 (\hbar c)^3} m\eta \,,
\label{etalink}
\end{eqnarray}
where the second equation follows from the thermodynamical formula for the number density of photons, $\zeta (3) = 1.20206$, and $mc^2 \simeq 0.94\,$GeV is the rest mass of a baryon.  This equation relating $\eta$ to the CMB temperature and baryon density is the key link to predicting the CMB temperature based upon BBN: 
knowing the value of $\eta$ and the baryon density, allows one to predict the temperature.

Today, the temperature is measured very precisely, $T_0 = 2.7255 \pm 0.0006\,$K \cite{Fixsen}, and the number density of photons is $411\,{\rm cm}^{-3}$.  The baryon density is
\begin{eqnarray}
    \rho_B & = &6.82 \eta_{10} \times 10^{-32}\,{\rm g\,cm}^{-3} \simeq 4.1\times 10^{-31}\, {\rm g\,cm}^{-3} \nonumber \\ 
   \Omega_B h^2 & = & 3.63\eta_{10}\times 10^{-3}  \simeq 0.022
\end{eqnarray}
where $\eta_{10} \equiv \eta /10^{-10}$, $\rho_{crit} = 3H_0^2/8\pi G = 1.88h^2\times 10^{-29}\,{\rm g\,cm}^{-3}$, and the Hubble constant $H_0 \equiv 100h\,$km/s/Mpc.  We know today, from both the CMB and BBN, that $\Omega_Bh^2 \simeq 0.022$ so that $\eta_{10}\simeq 6.1$.

The baryon-to-photon ratio also determines when the Universe transitioned from being radiation dominated, $\rho \simeq \rho_r$ to being matter dominated, $\rho \simeq \rho_B$.  This occurs at temperature
\begin{eqnarray}
    T_{EQ} = {30 \zeta(3) \eta m c^2\over k\pi^4} = 404 \eta_{10}\, {\rm K} \, ,
\end{eqnarray}
where $\rho_r = \rho_B$ at $T=T_{EQ}$.
This is an important milestone, because the small density perturbations that seed the formation of structure in the Universe can only grow due to gravitational instability when the Universe is matter-dominated.\footnote{In a universe like ours, with dark matter in addition to baryons, the formula for $T_{EQ}$ is modified by a factor of $\rho_{matter}/\rho_B \simeq 6$, where $\rho_{matter} = \rho_B + \rho_{DM}$.}

One last thing: the origin of this important number $\eta$.  Prior to 1980, the discussion about its origin focused on how to create the tremendous entropy implied by $\eta$.  Starting around 1980, $\eta$ was ``read'' another way, as  a measure of the tiny asymmetry between matter and antimatter.  During the earliest moments when, $kT \gg mc^2 \simeq 1\,$GeV ($t\ll 10^{-6}\,$sec), baryons and antibaryons were as abundant as photons, and $\eta$ quantified the slight excess of matter over antimatter that would ensure the existence of matter today after all of the antibaryons and most of the baryons annihilated.  Since then, the focus has been on the origin of this tiny excess due to out-of-equilibrium $B$, $C$, and $CP$ violating interactions in the very early Universe, known as baryogenesis \cite{baryogenesis}.  It now seems likely the high entropy is explained by the smallness of the excess of matter over antimatter, in turn controlled by $C$ and $CP$ violation.

\subsection{BBN clarified}

Three crucial events occur during the BBN epoch ($kT \sim 1\,{\rm MeV} - 10\,{\rm keV}$) and determine the outcome.  Their interplay and dependence upon $\eta$ are shown in Fig.~\ref{keyrates}.

The first event is the freeze out of the weak interactions
- the so-called $\beta$-reactions - 
that interconvert neutrons and protons and control their ratio:  $p + e^- \leftrightarrow n + \nu_e$, $n + e^+ \leftrightarrow p + \bar{\nu}_e$ and $\leftrightarrow p + e^- +\bar{\nu}_e$.  At a temperature $kT\simeq 0.7\,$MeV the neutron-to-proton ratio freezes in at a value of about $1/7$, rather than exponentially decreasing to zero as the lower mass proton state becomes thermodynamically favored.\footnote{It began slightly larger and slowly decreased to this value due to neutron decays.}  

Having free neutrons present - something that $\alpha\beta\gamma$ realized - is critical because means the first step of any nucleosynthesis is an EM interaction, $ n + p \rightarrow d + \gamma$, not a weak interaction as mentioned earlier.  The freeze out of the neutron-to-proton ratio is largely independent of $\eta$.

The second important event is nuclei becoming thermodynamically favored over free nucleons. A simple calculation of Nuclear Statistical Equilibrium (NSE) shows that nuclei become favored at a temperature \cite{KolbTurner}:
\begin{eqnarray}
    kT_{\rm nuclei} & \simeq & { B_A/(A-1) \over \ln \eta^{-1} + 1.5 \ln (mc^2 /kT_{\rm nuclei}) } 
     \simeq  0.25\,{\rm MeV} [1 + \ln \eta_{10} /35 ],
    \label{T_nuclei}
\end{eqnarray}
where $B_A$ is the binding energy of nucleus of atomic mass $A$, here taken to be that of $^4$He, or about 7\,MeV per nucleon.  Because of the very large number of photons per baryon, nuclei are not thermodynamically favored until a temperature much, much less than a typical nuclear binding energy, cf. the $\ln \eta^{-1}$ in the denominator of Eq.~\ref{T_nuclei}.  The temperature when nucleons are favored does depend upon $\eta$, and smaller $\eta$ lowers the temperature at which nuclei become favored and delays BBN.

The third crucial event is the freeze out of charged nuclear nuclear reactions due to Coulomb barriers.  Any nucleosynthesis beyond deuterium must involve charged-particle on charged-particle reactions, e.g., $d + d \rightarrow  {^4{\rm He}} + \gamma$ or $t + D \rightarrow n\ + \ $$^4$He, and because of the relatively low temperatures, Coulomb barriers - the electrostatic repulsion between charged nuclei - are important.  They suppress nuclear reaction rates by an exponential factor; for d(d,$\gamma$)$^4$He, that factor is approximately $\exp [-2(kT/{\rm MeV})^{-1/3} ]$.  

The nuclear reactions that are needed to produce anything beyond deuterium  cease occurring (``freeze out") at a temperature that can be estimated by comparing the expansion rate to the nuclear reaction rate (see \S\ref{secA1} or Ref.~\cite{KolbTurner} for more details):
\begin{eqnarray}
    kT_{\rm Coulomb} \sim {0.03\,{\rm MeV} / [1+ \ln \eta_{10} /7]^3 }
\end{eqnarray}
In this case, smaller $\eta$ means nuclear reactions freeze out at a higher temperature and end BBN earlier.  

\subsection{Summary}

Bringing it all together, nucleosynthesis beyond deuterium requires that nuclei become thermodynamically favored before nuclear reactions freeze out due to the effect of Coulomb barriers , or $kT_{\rm nuclei} \ge T_{\rm Coulomb}$.  This corresponds to $\eta > 10^{-11}$.  As seen in Fig.~\ref{keyrates}, this puts a lower bound to $\eta$ based upon  nucleosynthesis taking place.

For $\eta > 10^{-11}$, essentially all the neutrons end up in the thermodynamically-favored $^4$He nuclei, around 10\% to 30\% by mass, depending logarithmically upon $\eta$. By the time other nuclei become more thermodynamically favored than $^4$He, Coulomb barriers have shut off nuclear reactions, unless $\eta$ is exceedingly large, $\eta \gg 10^{-3}$.  

It follows from this basic picture that when big-bang nucleosynthesis takes place ($\eta > 10^{-11}$):
\begin{enumerate}
\item{} A significant amount of $^4$He (10\% to 30\%) is  produced, whose amount depends logarithmically upon $\eta$ and is limited by the neutron supply 
\item{} A large amount of hydrogen remains unburnt (70\% to 90\% by mass) because nucleosynthesis is limited by the available neutrons 
\item{} Little nucleosynthesis beyond $^4$He occurs because the binding energy of $^4$He is so large and other nuclei are not thermodynamically favored until Coulomb barriers have already quenched further nucleosynthesis
\item{} Some deuterium remains unburnt, with an abundance that decreases rapidly with $\eta$ since more deuterium is used up when the baryon density is larger
\item A small amount of $^3$He and an even smaller amount of $^7$Li is made
\end{enumerate}
For $\eta < 10^{-11}$, only deuterium is synthesized (as much $D/H \sim 10^{-2}$).  For the smallest values, $\eta \ll 10^{-11}$, the amount of deuterium produced falls as $\eta$ because its production is rate limited by small abundance of nucleons, $\Gamma_{p + n \rightarrow d + \gamma} < H$ for $\eta < 3\times 10^{-13}$.

\begin{figure}[h]
 \centering
\includegraphics[width = 0.85\textwidth]{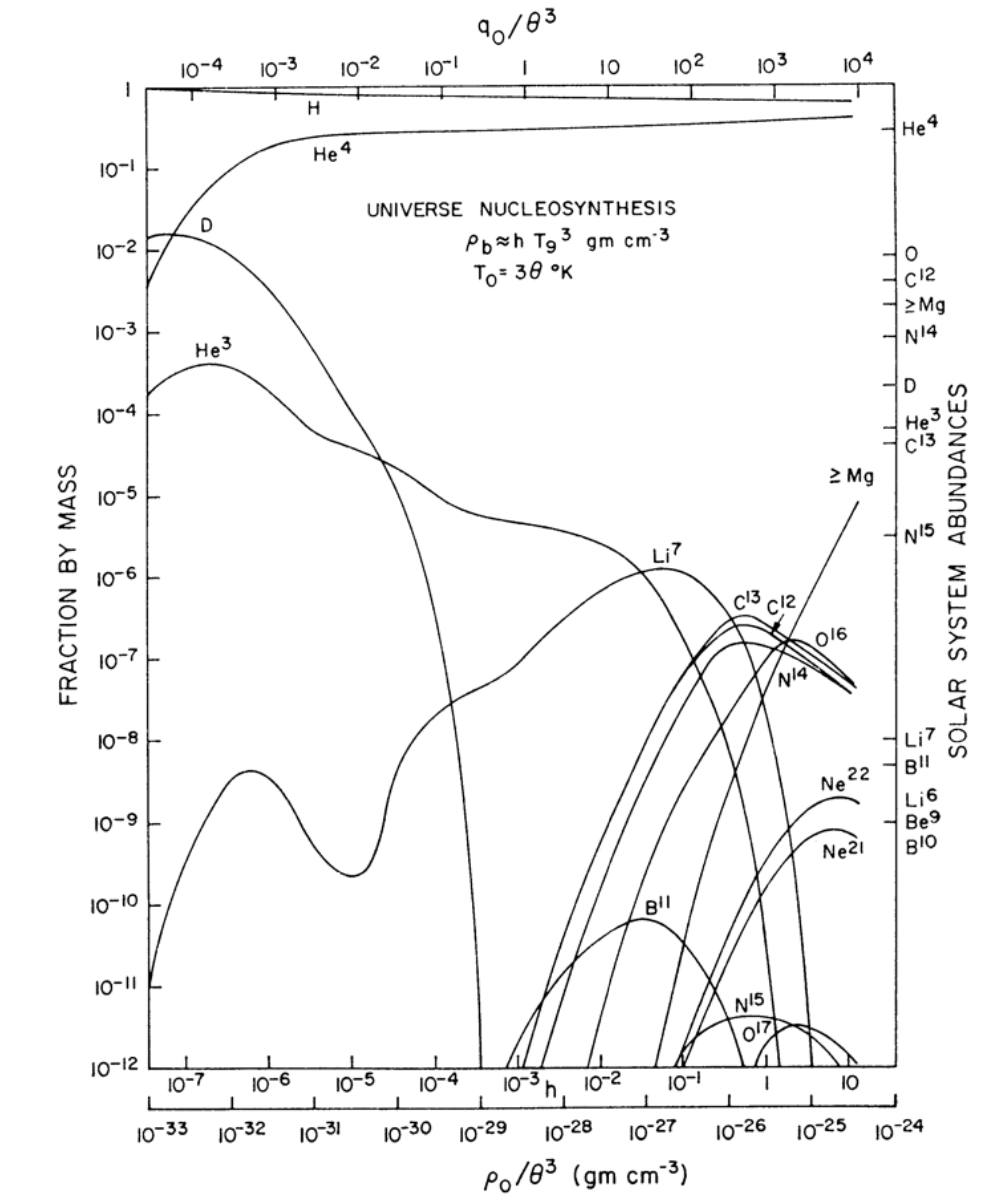}
\caption{Abundances computed by Wagoner, Fowler and Hoyle \cite{WFH}; cross sections and physics has since been updated, but the results are qualitatively correct, and even quantitatively when viewed on a log-log plot. Note:  their ``little $h$" is not $H_0/100\,$km/s/Mpc; it is their version of the baryon-to-photon ratio, $h_{WFH} \equiv (\rho_B/{\rm g\,cm}^{-3})/(T/10^9\,{\rm K})^3 = 3.37 \times 10^{4}\eta$. 
The WFH calculations span nine orders-of-magnitude, $\eta \simeq 10^{-12}$ to $10^{-3}$, a feat in itself.}
\label{BBN_abundances_eta}     
\end{figure}

The detailed calculations of Wagoner, Fowler and Hoyle \cite{WFH} (and many other calculations since) bear out this picture; see Fig.~\ref{BBN_abundances_eta}. Big bang nucleosynthesis is described in more detail in many places; see e.g., \cite{KolbTurner,Weinberg72}. 

\section{A brief history of BBN physics}
The fog at the frontiers of discovery is  always thick, and a clear understanding of a big, new idea with all the details in place usually takes awhile and the path has its twists and turns.  BBN was no different.  It was the early days of nuclear physics and nuclear data was sparse;  particle physics was not yet a field; scientific computing was in its infancy, and astronomy was done with photographic plates and visible light, with radio astronomy just beginning and still in the realm of physics.  And cosmology was in its infancy.

This section describes my reading and interpretation of the original papers.  It begins with the landmark $\alpha\beta\gamma$ paper in 1948, and is bookended some 20 years later with the equally important paper by Wagoner, Fowler and Hoyle \cite{WFH} in 1967.  Because of the initial focus on non-equilibrium nucleosynthesis, the story ends far from where it began:  BBN is a mostly equilibrium story with a couple of significant deviations from thermal equilibrium.

\subsection{Gamow's theory}

In Gamow's neutron-capture model, matter in the Universe begins entirely comprised of neutrons.  The protons needed for nucleosynthesis are produced later by neutron decays, setting the timescale, $t\sim \tau_n \sim 1000\,$sec.  The limited neutron-capture data were used to identify the nucleon density needed to move up the periodic table by successive neutron captures.  The temperature and nucleon density at $t\sim 1000\,$sec then fix the baryon-to-photon ratio, $\eta \sim 10^{-11}$.  

Alpher and Gamow published additional papers with numerical calculations of the production of deuterium and other light elements \cite{gamow48,alpherherman48,alpher48} as well as predictions for the temperature of the CMB.\footnote{All told, more than a dozen papers on BBN were published by Alpher and various co-authors, mostly with Robert Herman and only the first with George Gamow.  The ones directly cited are the most relevant for my discussion of the physics and the history.  A fuller list of the papers is given in Ref.~\cite{alpherpapers}.} All of these papers were predicated on an initial state of pure neutrons, with only neutron decays and captures taken into account.

Two  critiques of Gamow's theory were immediately raised.  In an unpublished correspondence with the authors, Fermi and Turkevich \cite{FT1950a} pointed out that the absence of long-lived isotopes with mass 5 and 8 prevents significant nucleosynthesis beyond $^4$He in the neutron capture model (all the nuclides of mass 5 have lifetimes in the yoctosecond range; and for mass 8 at most in the msec range).  This is a showstopper for the neutron-capture model, but is less relevant for the way in which BBN actually works (more in \S\ref{missed}).

Hayashi \cite{Hayashi50} made an even more important point:  weak interactions involving neutrons and protons - the $\beta$-reactions -
should occur rapidly because of the thermal bath of $e^\pm$ pairs and neutrinos and antineutrinos and regulate the ratio of neutrons-to-protons.\footnote{Unlike weak interactions among nuclei, the $\beta$-reactions are important because they involve nucleons interacting with the thermal photons, electron/positron pairs and neutrinos which are $\eta^{-1} \sim 10^{10}$ times more abundance than nucleons and thus occur at a much fast rate.}  This precludes the initial state of pure neutrons assumed in $\alpha\beta\gamma$, and makes the results of BBN insensitive to the initial value of the n/p ratio.  This led to a series of papers, including one by Alpher, Follin and Herman \cite{afh}, that focused on computing the neutron-to-proton ratio by numerical integration.  (While the discovery of the neutrino did not happen until 1956, Fermi's theory of the weak interactions was well developed and well known.)

The other key element on the path to the modern theory was to incorporate  a nuclear-reaction network that included charged-particle reactions as well as neutron captures.  That was done around the same time -- circa 1949 or 1950 -- in an {unpublished} paper by Fermi and Turkevich \cite{FT1950}, entitled, {\it Thermonuclear reactions in the first hour of an expanding Universe}.  While unpublished, it is described in detail in the 1950 {\it Rev. Mod. Phys.} article by Alpher and Herman \cite{AlpherRMP}.  The Fermi-Turkevich network contained 28 reactions, and they tracked the abundances of H, He, Li, and Be.\footnote{I had the pleasure of knowing Tony Turkevich as a colleague at UChicago and I even gave several talks about BBN which he attended.  He may have briefly mentioned his and Fermi's work, but I can't remember.  I was unsuccessful in finding their paper in Fermi's collected works in the UChicago library.  Turkevich's papers do not exist there.  However, his son Leonid (Lee) Turkevich, a classmate of mine at Stanford, told me a few years ago he has his father's unsorted papers.  He recalled his father telling him about the work with Fermi, being especially proud of their prediction of a large amount of $^4$He.  Lee also mentioned that his father and Gamow were life-long friends and that his father and mother drove from Chicago to Colorado to attend Gamow's funeral in 1968.} 

However, Fermi and Turkevich also assumed an all neutron initial state and did not include the $\beta$-reactions that set the neutron-to-proton ratio nor the inverse reactions that are so important because of the high entropy of the Universe.  Their predicted $^4$He mass fraction for $\eta \simeq 2\times 10^{-11}$ of 0.28 is about a factor of 2 high compared to the correct calculation, though a quantitative comparison  doesn't really make sense since their calculation leaves out so many important reactions. 
Nonetheless, they were on the path to correctly identifying that the outcome of BBN is largely $^4$He production and not the heavier elements.

Taken together, the Hayashi insight and the Fermi/Turkevich network if combined would take one almost all the way to WFH; however, the paper of Fermi and Turkevich was never published.\footnote{One can speculate why; among other things, Turkevich was deeply involved in work involving nuclear weapons test monitoring at Los Alamos during this period \cite{Turkevichmemoir}.  Perhaps, Fermi lost interest when he saw that he missed the $\beta$-reactions, based upon his theory, that control the neutron-to-proton ratio. Nonetheless, the calculations they carried out were state of the art for the time and used some of the most powerful computers at the time, e.g., the ENIAC and the MANIAC at Los Alamos.}  

\subsection{The dawning of the modern era}
In 1964, just before the discovery of the CMB, Hoyle and Tayler did put the two together in a paper entitled, {\it The mystery of the the cosmic Helium abundance} \cite{HoyleTayler}.  Here they combined approximate tracking of the $n/p$ ratio and a small nuclear network (not described in detail) to predict a $^4$He abundance of around 36\%, which they concluded was consistent with estimates of the observed He abundance, whose value was not well determined.  Equally important, they made the case for a large primordial $^4$He abundance that could not be simply explained by stars but could be explained by BBN.\footnote{They also noted that the cosmic Helium problem could be solved if similar conditions were achieved in a supermassive star.  This is a theme that would continue in the WFH paper and presumably had to do with Hoyle's disdain for the hot big-bang model.}

Hoyle and Tayler 
also state in the penultimate paragraph on page 1109, ``we can now say that if the Universe originated in a singular way the He/H ratio cannot be less than about 0.14,''  which corresponds to a $^4$He mass fraction of about 36\%.\footnote{Various authors track different quantities; under the (good) assumption that all the neutrons wind up in $^4$He, $Y = 2(n/p)/[1+ (n/p)] = 4(n_{He}/n_H)/[1+4(n_{He}/n_H)]$, where the Helium-to-Hydrogen ratio $n_{He}/n_H = (n_n/2)/(n_p - n_n)$.}  Their statement is based upon the two paragraphs above, in which they argue that any variations in the cosmological conditions and parameters would only increase $^4$He production.  Of course, we know that for $\eta < 10^{-11}$ there is essentially no nucleosynthesis and only at very high values of $\eta$ does $^4$He production reach 36\%.  %%Moreover, the primordial $^4$He abundance is only about 25\%. 
There is not enough detail in the paper to see where they went wrong, but we do know that Hoyle was not a fan of the big-bang theory, having invented its moniker in order to refer to it derisively.

After the discovery of the CMB, Peebles \cite{PJEPBBN} carried out a similar calculation, with a limited reaction network and an approximate treatment of the $n/p$ ratio, arriving at a $^4$He abundance of between 27\% and 30\% (for $\eta = 4 \times 10^{-10}$).  Both Peebles and Hoyle and Tayler halted their numerical integrations at a temperature of around $5\times 10^9\,$K, without a clear argument for doing so.  Unlike, the Hoyle-Tayler paper, Peebles' paper does show the correct qualitative variation of $^4$He production with baryon density (or $\eta$) in his Fig.~1.\footnote{While Peebles cites the Hoyle-Tayler paper for its discussion of the evidence for a large primordial $^4$He abundance, it does not mention their similar BBN calculations or compare his results with theirs.}

Around the same time, Wagoner, Fowler and Hoyle \cite{WFH} put it all together in their landmark paper.  They tracked the abundances of nuclides $A\le 23$  individually, and for $A\ge 24$ collectively; they included the thermal bath of photons, neutrinos, and electron/positron pairs, and numerically followed the n/p ratio using no approximations; their nuclear-reaction network contained 79 reactions as well as the most up to date nuclear data; and their numerical integration began at $T \sim 10^{11}\,$K and continued down to $T \sim 5\times 10^7\,$K, until well after Coulomb barriers had quenched all nuclear reactions.  The results are shown in Fig.\ref{BBN_abundances_eta}.\footnote{Peebles' paper was submitted for publication in February 1966 and appeared in March 1996 \cite{PJEPBBN}; WFH gave a talk at the National Academy of Sciences meeting in April 1996, presenting their preliminary results \cite{WFH_NAS_1966}.  Their 47-page paper was submitted in September 1966 and published in April 1967.}

\begin{figure}
\centering
  \includegraphics[width=0.75\linewidth]{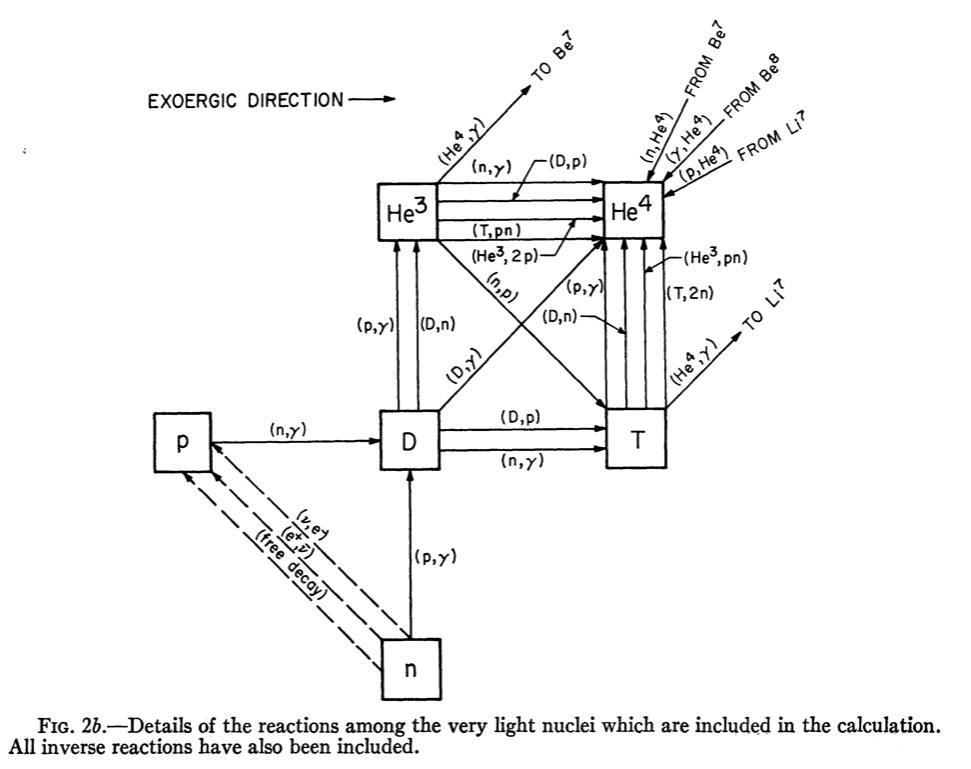}
  \caption{The smaller WFH network of nuclear reactions that leads to the production of $^4$He \cite{WFH}.} \label{fig:light-network}
\end{figure}

\begin{figure}
\centering
  \includegraphics[width=0.9\linewidth]{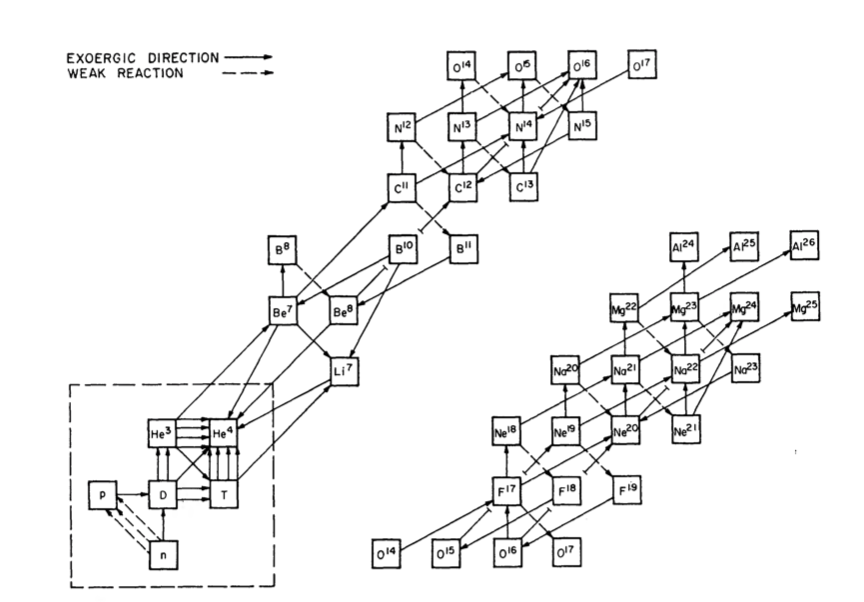}
  \caption{The complete WFH network of 79 BBN nuclear reactions \cite{WFH}.} \label{fig:network}
\end{figure}

Since then, BBN calculations have been updated for changes in nuclear-reaction rates (most importantly, the neutron lifetime) and percent-level physics effects have been accounted for to obtain greater precision (e.g., finite nucleon mass, radiative corrections, finite temperature quantum corrections, and the slight heating of the neutrinos by $e^\pm$ annihilations).  Nevertheless, WFH is the paper that included all the physics and marks the beginning of the modern age of big-bang nucleosynthesis.

\subsection{The modern theory}

The modern picture of BBN is a largely a story of thermal equilibrium, with the high entropy preventing significant nucleosynthesis until a temperature of about $kT \simeq 0.2\,$MeV, when nuclei became thermodynamically favored.  The  neutrons needed for BBN arise because the weak interactions that regulate the $n/p$ ratio cease to be effective around 1\,sec/1\,MeV and n/p freezes in at a value of 1/7.  
Without the freeze out of the weak interactions there would be no neutrons left by the time nuclei became thermodynamically favored and hence no BBN.  The falling temperature and the exponentially rising effect of Coulomb barriers ends BBN at around 200\,sec/$kT \simeq 0.05\,$MeV, shortly after $^4$He is statistically favored (provided that $\eta > 10^{-11}$).  

\begin{figure}[h]
\begin{subfigure}{0.40\textwidth}
\includegraphics[width=1.0\linewidth]{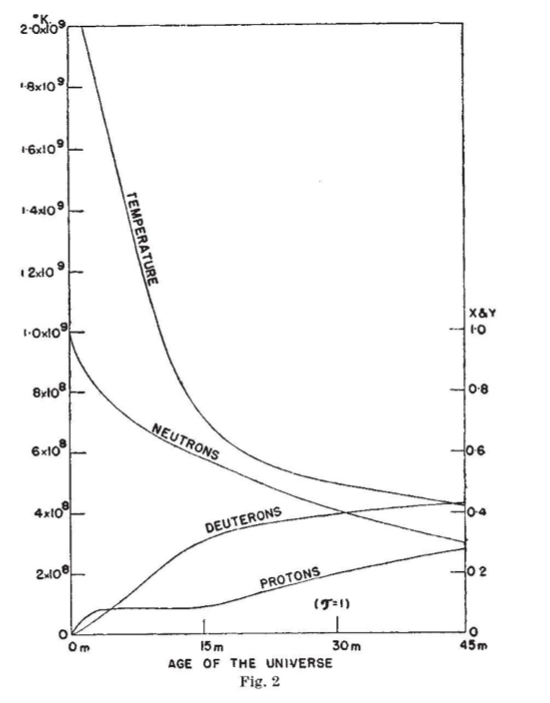}
\end{subfigure}
\begin{subfigure}{0.59\textwidth}
\includegraphics[width=1.0\linewidth]{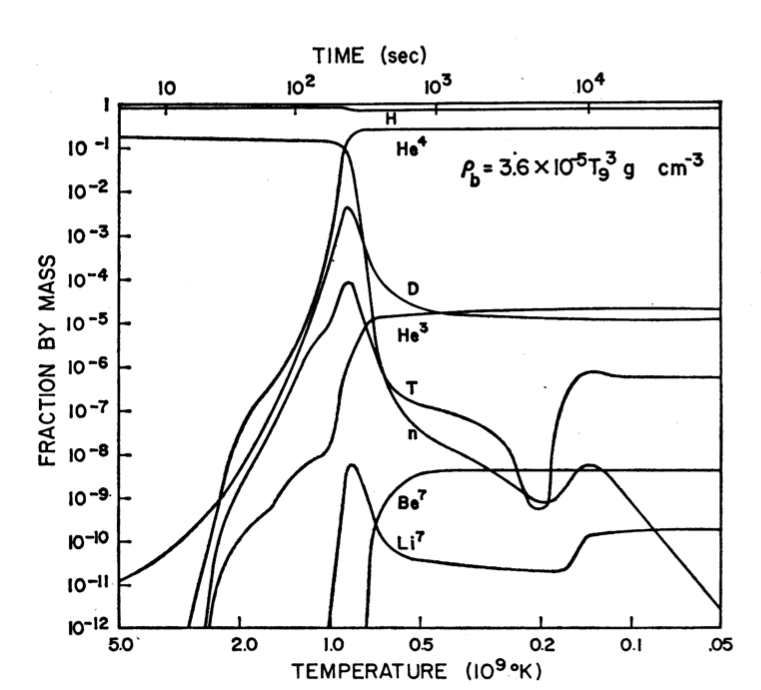}
\end{subfigure}
\caption{Left:  Time evolution in the neutron-capture model \cite{gamow48} (left), and in the modern theory of BBN \cite{WFH} (right).  The factor of 10 extra time in the neutron-capture model is for neutrons to decay and produce the protons needed for nucleosynthesis.  In the modern theory, the neutrons are present because of the freeze out of the $\beta$-reactions (at around 1\,sec); nucleosynthesis is over by 200\,sec because Coulomb barriers prelulde any further nuclear reactions.}
\label{BBNevolution}
\end{figure}

The ``baking time'' sums up the difference between Gamow's theory and the modern theory:  In the neutron-capture model the needed protons are produced by neutron decays which takes almost an hour, and in the modern theory nucleosynthesis must occur before the neutrons decay and before Coulomb barriers become insurmountable and takes only 3 minutes; cf., Fig.~\ref{BBNevolution}.\footnote{A theorist can't resist asking if there is any scenario in which the neutron-capture model would be relevant.  It might be a good approximation for a universe that began with pure neutrons at an age of around 500\,sec and temperature of about $0.05\,$MeV.  In this case, the $\beta$-reactions would be ineffective, and inverse reactions with photons that break up nuclei would be as well.  Thus both could be neglected, and the calculations of Fermi and Turkevich might be a good approximation.  Unfortunately, the computed $^4$He mass fraction is only 20\% (too low) and the deuterium abundance is 1\% (too high); see Fig.~20 of Ref.~\cite{AlpherRMP}.} That is why Steven Weinberg's well-known popular book on cosmology is entitled, {\it The First Three Minutes} \cite{Weinberg}, and not, as the Fermi/Turkevich paper is entitled, {\it ... the first hour of the Universe}!

\section{Some misses}\label{missed}
\subsection{Two misunderstandings}
The complicated path from the neutron-capture model that aspired to synthesize the entire periodic table of elements to the present mostly equilibrium model of the synthesis of the light elements has left its imprint on the discussion of big-bang nucleosynthesis today.  Statements that the deuterium bottleneck delaying the synthesis of $^4$He and the absence of long-lived nuclei with mass 5 and 8 preventing nucleosynthesis beyond Helium, neither of which is true, appear in review articles \cite{BBNreviews}, Wikipedia's discussion of BBN and in the many on-line lectures about BBN that can be found through Google searches.  In reality, the real culprits are the high entropy and Coulomb barriers.  

Recall, Fermi and Turkevich pointed out that the absence of stable nuclei of mass 5 and mass 8 is a roadblock to the neutron-capture model going beyond $^4$He synthesis.  This is correct - for the neutron-capture model.  However, in the modern theory, charged-particle reactions  bridge the gap; e.g., $^3$He + $^4$He $\rightarrow$ $^7$Be + $\gamma$ and so on; see Figs.~\ref{fig:light-network} and \ref{fig:network}. 
Nonetheless, one still sees the lack of significant nucleosynthesis beyond $^4$He blamed on the absence of stable nuclei with mass 5 and 8.  The real reason is simply thermodynamics:  by the time heavier nuclei are thermodynamically favored, Coulomb barriers have shut off any further nucleosynthesis.  It is a simple matter to add a stable mass 5 and/or 8 nuclide to the BBN code and find that nothing changes!

Blaming the late onset of BBN on the small binding energy of deuterium (about 1.1\,MeV per nucleon) is another misunderstanding.  As discussed earlier, the cause is really the high-entropy of the Universe, which means nuclei do not become thermodynamically favored until $kT_{\rm nuclei} \simeq 0.25\,$MeV.  

The term itself appears to have been coined by Weinberg in his discussion of BBN (see p.~553 of Ref.~\cite{Weinberg72}), where he says, ``The low binding energy of deuterium thus serves as a bottleneck, which delays the formation of complex nuclei ...''.  His words are not entirely incorrect, but they are very misleading.  Consideration of NSE alone would have the $^4$He abundance soar at a temperature of about $0.25\,$MeV, far below the binding energy of 7\,MeV/nucleon; it actually occurs closer to $0.1\,$MeV.  The delay is a ``supply chain" issue, with multiple parties to blame.

All of the reactions that produce $^4$He have rates that are proportional to  $n_B X_D \left< \sigma v \right> \propto \eta T^3X_D \left< \sigma v \right>$ ($X_D \equiv n_D/n_B$).  Each one of the three factors contributes to the delay in converting all of the available neutrons into $^4$He when $^4$He becomes thermodynamically favored:  the small baryon-to-photon ratio $\eta$, the fraction of nucleons that are deuterons, $X_D \sim 10(kT/mc^2)^{3/2} \exp (2.22\,{\rm MeV}/kT )$, and the thermally-averaged reaction rates which are suppressed by Coulomb barriers.  To blame it on deuterium alone is to neglect the other factors at play. An even more spirited defense of deuterium and longer discussion of both misunderstandings is given in Chapter 4 of Ref.~\cite{KolbTurner}.

\subsection{A missed opportunity}

The story of the discovery of the CMB and the missteps along the way is  fascinating and well documented \cite{PJEPCC,Weinberg,PPP}.  On the theory side, there was the opportunity of a lifetime -- predicting the existence of the CMB and its temperature on the basis of BBN.  Here I focus on the predictions that {\it could} have been made and the predictions that {\it were} actually made, remembering that 20/20 hindsight is always much clearer.

The formula that guides this discussion is Eq.~\ref{etalink}, which relates $\eta$, the baryon density today and the CMB temperature today, or at any epoch.  The desired outcome for BBN determines $\eta$, and then, given the baryon density, the CMB temperature is predicted:
\begin{eqnarray}
    T_{\rm CMB} & = & \left[ {\rho_B/(6.86 \times 10^{-32}\,{\rm g\,cm}^{-3}) \over \eta_{10}} \right]^{1/3} 2.73\,{\rm K}  \\     
  & = & \left[ {274 \Omega_B h^2 \over \eta_{10} } \right]^{1/3} {2.73}\,{\rm K} \nonumber 
\label{predicted_T}
\end{eqnarray}
where the two variants allow the baryon density to be given either in ${\rm g\,cm}^{-3}$ or as fraction of critical density times the Hubble constant squared.  Note too that the predicted temperature is relatively insensitive to baryon density and $\eta$, because of the cube root.

\subsection{The counterfactual}\label{counterfactual}

Without detailed calculations, based solely upon the basic physics described in \S2, one could say $\eta$ must be greater than $10^{-11}$ to have significant nucleosynthesis, i.e., $^4$He mass fraction of up to 30\% or so and some leftover D.  With a calculation like that of Hoyle and Tayler \cite{HoyleTayler}, one could go further than that and say $\eta \sim 10^{-10} - 10^{-9}$ is required to produce the observed primordial $^4$He mass fraction of 20\% to 30\%.

To predict the CMB temperature, one needs an estimate of the baryon density.  Today, we know it very precisely, $\Omega_B h^2 = 0.022$ or $4.1\times 10^{-31}\,{\rm g\,cm}^{-3}$.  Pre-CMB discovery, circa 1960, much less was known.  According to Peebles \cite{PJEPBBN}, based upon the mass observed in galaxies, $\rho_B \simeq 7\times 10^{-31}\,{\rm g\,cm}^{-3}$.  A similar estimate was made by Zel'dovich (more below), $\rho_B \simeq 3\times 10^{-31}\,{\rm g\,cm}^{-3}$\footnote{Both of these estimates are greater than our better knowledge of the baryon density today.  Why?  The story is not simple, but they are based upon using the measured mass-to-light ratios of nearby galaxies, estimated to be around 30 in solar units, and converting that into a mass density by multiplying by the luminosity density \cite{Sidney1961}, which introduces a linear dependence upon the Hubble constant, which at the time was almost a factor of two larger.  In addition, their mass-to-light ratios appear to include some of the dark matter which is non-baryonic.}.  Using the mean of these two numbers, the CMB predictions are:
\begin{eqnarray}
     T_{\rm CMB} & < & 11\,{\rm K} \qquad \qquad \ \, {\rm some\ nucleosynthesis:\ }\eta >10^{-11}  \\
    T_{\rm CMB} & = & 2.5 - 5.3\, {\rm K} \ \ \ \ \  ^4{\rm He\ synthesis:\ }\eta = 10^{-10}-10^{-9} 
\end{eqnarray}

At the time, the Hubble constant was pegged at around 100\,km/sec/Mpc (or $h =1$), so the critical density would correspond to $\rho_B \simeq 2\times 10^{-29}\,{\rm g\,cm}^{-3}$, about 40 times larger than estimates for the matter density based upon the mass seen in galaxies.  While it would be an unlikely choice, with this extreme assumption for the baryon density, the predictions for the CMB temperature are:
\begin{eqnarray}
     T_{\rm CMB} & < & 38\,{\rm K} \qquad \qquad \  {\rm some\ nucleosynthesis:\ }\eta >10^{-11}  \\
    T_{\rm CMB} & = & 8.2 - 18 \, {\rm K} \qquad  ^4{\rm He\ synthesis:\ }\eta = 10^{-10}-10^{-9} 
\end{eqnarray}  
Any of these predictions could have been made, and the discovery of the CMB with a temperature of around 3\,K would have been a real coup for the theorist who did!

\subsection{The reality}
Now for what actually transpired before the discovery.  Beginning with the least controversial, Hoyle and Tayler, and then, Zel'dovich and the Soviet school, and finishing with the most controversial, Alpher and Herman.

\subsection{Hoyle and Tayler}
Prior to the discovery of the CMB, Hoyle and Tayler \cite{HoyleTayler} carried out an almost-modern calculation of BBN, and further, they assembled the case for a large, primordial amount of $^4$He.  However, they made no prediction for the temperature of the CMB.  Two other things worked against them making a prediction:  first, their calculations overestimated the amount of $^4$He production (because of the lack of detail, it is hard to figure out the source of this error); and second, they did not explore the dependence of the $^4$He production on the present temperature.  And of course, Hoyle was not a fan of the big-bang theory, and saw BBN as a means of falsifying it.

\subsubsection{Zel'dovich}
In discussing the history of BBN, I did not mention Zel'dovich and the Soviet school of cosmology.  In part, because much of the work in the Soviet Union found its way into western science slowly if at all. Zel'dovich wrote a comprehensive review of cosmology just {\it before} the discovery of the CMB \cite{Zel64}.  In it, there is a remarkable paragraph on page 491, which discusses BBN. While it lacks equations or any detail, it is fully consistent with the picture I have discussed here and makes it clear that he had zeroed in on the correct physics.  In that paragraph, Zel'dovich describes two possibilities, a low-entropy universe and a high-entropy universe, and dismisses them both!\footnote{That paragraph is expanded upon in a paper published a year later by Smirnov \cite{Smirnov}, where Smirnov describes calculations that he carried out at the behest of Zel'dovich and which underpin the conclusions in the paragraph.  Smirnov discusses the context of his work in Ref.~\cite{PPP}, pages 92-99.}

For the low-entropy universe, $\eta \sim 10^{-8}$, the $^4$He production is 30\%, which Zel'dovich concluded exceeds that seen in old stars.  For  the baryon density he uses, $\rho_B = 3\times 10^{-31}\,{\rm g\,cm}^{-3}$, this corresponds to a CMB temperature of 1\,K.  For the high-entropy universe, $\eta \sim 10^{-12} $,  he states that $^4$He production is less than 10\% (also true) and the CMB temperature today would be $20\,$K.  He goes on to say that the Universe would still be radiation-dominated which would prevent structure formation (also true), and argues against the high-entropy case as well.

Instead of considering an intermediate case -- say $T_{\rm CMB} \simeq 3\,$K -- where the $^4$He production would take on the Goldilocks value, Zel'dovich dismisses BBN and goes on to describe his favorite model, the cold big-bang. A missed opportunity!  Two years later, after the discovery of the CMB, he writes another review of cosmology \cite{Zel67}, quickly and enthusiastically embracing the hot big bang model and discussing all of its consequences including BBN.  

\subsubsection{Alpher and Herman}
This brings us to the most controversial, best-publicized and very confusing prediction of the CMB temperature, that made by Alpher and Herman in 1948 \cite{alpherherman48}.  Their paper is a short note to {\it Nature}, correcting some numerical errors in the paper published by Gamow in {\it Nature} two weeks earlier \cite{gamow48}.  Gamow's paper itself is based calculations from Alpher's PhD thesis \cite{alpher48};  and, he adds a new element to the discussion:  the formation of galaxies.  He notes that for his favored parameters ($\eta = 6\times 10^{-11}$), when the Universe became matter-dominated and structure could begin to grow, the Jeans mass is comparable to that of small galaxies.\footnote{Gamow was unaware of Lifshitz's pioneering and more extensive work on the evolution of small density perturbations and the growth of structure \cite{Lifshitz}.}

While Gamow's paper could have predicted a CMB temperature, it did not.  After correcting Gamow's numerical errors and near the end of their paper, Alpher and Herman predict a present temperature of 5\,K for the relic radiation, remarkably close to the actual value.  While Alpher often touted this prediction (see e.g., \cite{AlpherHermanPT}), as I now explain, their prediction is just a happy coincidence.  

Based upon Alpher's numerical calculations of the neutron-capture model, Alpher and Herman chose a matter-to-radiation ratio that corresponds to $\eta = 4 \times 10^{-12}$.\footnote{I have translated Alpher's baryon density of $4.8 \times 10^{-4}\,{\rm g\,cm}^{-3}$ at a photon temperature of $1.5\times 10^{10}\,$K into this value for $\eta$.}  
This is a very high-entropy Universe, where essentially no nucleosynthesis would take place.  Further, such a universe would still be radiation-dominated today and for any reasonable estimate of the matter density, the CMB temperature would be large today.  This is essentially the high-entropy model Zel'dovich correctly dismissed.  

However, recognizing a radiation-dominated Universe today would be problematic, Alpher and Herman go on to invoke large negative-curvature to solve the problem and find a present temperature of $T_{\rm CMB} = 5\,$K.  Close to the right answer, but fortuitously so.  

Why do I say fortuitously so?  It is based upon a model of nucleosynthesis -- neutron-capture -- that is incorrect!  
And there is more wrong.  For their model, the Universe enters a negative-curvature dominated phase, where the scale factor increases linearly with time, at an age of about $10^8\,$yrs.  At the present age of the Universe, which they assume about 2\,Gyr, the baryon density would have fallen to a very low value, about $1.8\times 10^{-32}\,{\rm g\,cm}^{-3}$, a factor of 20 smaller than estimates of the matter density at the time.  Worse yet, their model was never matter-dominated, transitioning directly from being radiation-dominated to curvature-dominated.  Lifshitz had already shown that small density inhomogeneities only grow in a matter-dominated Universe, and so the structure we see today would never have formed \cite{Lifshitz}. More details about how they arrived at this cosmological model and their prediction are given in \S\ref{secAH}.

\subsection{More twists}
Finally, I mention two more twists. An observational upper limit to $T_{\rm CMB}$ was already known before the discovery of the CMB, at least to Hoyle.  In 1940, McKellar's \cite{McKellar} measurements of the population levels of interstellar CN and CH radicals constrained any ``universal radiation field," to have a temperature $T_{\rm CMB} < 5\,$K.  In fact, Hoyle used McKellar's result as an argument against Gamow's hot big bang model in debates the two engaged in.  After the discovery of the CMB by Penzias and Wilson, the population levels of interstellar CN were used to make important determinations of the CMB spectrum at shorter wavelengths \cite{PPP}, showing that it was consistent with a blackbody and not a Rayleigh-Jeans spectrum.

The second twist, involves the 1963 paper by Doroshkevich and Novikov \cite{DN1963}, entitled {\it Mean density of radiation in the Metagalaxy and certain
problems in relativistic cosmology}.  In it, they estimate the spectral distribution of radiation from all the galaxies in the Universe and briefly discuss the prospects for checking the ``Gamow'' theory, showing that a 1\,K blackbody rises above the radiation expected from galaxies in the radio portion of the spectrum.  According to Novikov and Doroshkevich \cite{Novikov_PPP}, their work was motivated by Zel'dovich and his Soviet group thinking about the consequences of a hot beginning.  The paper went unnoticed in the west.

\subsubsection{The long road to BBN}

It took a village and some 20 years to put the modern picture of BBN  together.  It  began in 1948 with Gamow's neutron-capture theory \cite{abc}, which introduced the radiation-dominated Universe and nuclear physics to cosmology.  In 1950, using Fermi's theory of the weak interactions, Hayashi added the $\beta$-reactions that control the neutron-to-proton ratio \cite{Hayashi50}.  Fermi and Turkevich \cite{FT1950} introduced the nuclear reaction network needed to track nucleosynthesis.

In 1953, Alpher, Follin and Herman \cite{afh} set up the modern calculation for the neutron-to-proton ratio, whose frozen in value determines the production of $^4$He.  In 1964, Hoyle and Tayler \cite{HoyleTayler}, building upon the work of Hayashi, Fermi and Turkevich, and Alpher, Follin and Herman, carried out the first numerical calculation for the predicted yield of $^4$He, and made the case for the existence of a primordial abundance of $^4$He of around 30\%.

In 1965 the CMB was accidentally discovered by Penzias and Wilson \cite{PW}, and in 1966 Peebles \cite{PJEPBBN} carries out a more refined version of the Hoyle and Tayler calculation.  Just months later, Wagoner, Fowler and Hoyle published their landmark paper on big-bang nucleosynthesis, the forerunner of all modern BBN calculations.

\section{Concluding remarks}
\subsection{Legacy}
Big-bang nucleosynthesis is a cornerstone of modern cosmology.  Its correct prediction of a large primordial abundance of $^4$He was one of the first successes of the hot big-bang model.  More recently, the measurement of the primordial deuterium abundance was used to pin down the baryon density to sub-percent-level precision, $\Omega_B h^2 = 0.02166 \pm 0.00015$ \cite{Cooketal}.  This determination is consistent with a determination based upon measurements of the CMB made by the Planck Satellite, $\Omega_B h^2 = 0.02237 \pm 0.00015$ \cite{PlanckLegacy}.  

This is an impressive consistency check of today's cosmological paradigm $\Lambda$CDM \cite{TurnerPNAS}: Two determinations of the baryon density, one based upon nuclear physics that took place seconds after the big bang and the other based upon the physics of the CMB, some 400,000 yrs later, agree to sub-percent precision.  

This measurement of the baryon density together with a similarly precise determination of the {\it total} matter density, which is almost 6 times larger and more than 50$\sigma$ different, $\Omega_M h^2 = 0.143 \pm 0.0011$, provides the linchpin in the case for non-baryonic dark matter.  Particle dark matter is a central pillar of $\Lambda$CDM, and the best candidate is a yet-to-be-discovered, long-lived particle.

Because BBN is so well understood and its predictions so precise, since 1977 it has been used as a probe of particle physics, e.g., by constraining the numbers and kinds of light particle species present in the thermal bath around the time of BBN, $kT \sim 1\,$MeV \cite{SSG}.

The legacy of BBN also includes the r-process in nuclear astrophysics, the building up of the heaviest elements through rapid neutron capture - sound familiar?  Since the ambitions of $\alpha\beta\gamma$ to synthesize all the chemical elements in the big bang were not realized, astrophysicists turned to exploding stars for the non-equilibrium conditions needed to make the heaviest elements through rapid neutron capture \cite{Cameron57,BBFH57}.\footnote{While it has been widely said that Hoyle was spurred by Gamow's claim to be able make all the elements in the big bang to think more broadly about stellar nucleosynthesis, at least one source disputes this \cite{Croswell}.}

The grandest legacy of BBN is the fact that most of the baryons in the Universe were left unburnt and with them a free energy of about 8.5\,MeV per nucleon.  It is this free energy that literally supplies all the energy on Earth and most of the energy in the Universe.  Stars fuse the light elements releasing the sunlight that heats Earth, drives the winds and ocean currents, and is harvested by solar arrays.  Nuclear fission processes inside Earth also heat it and in nuclear reactors produce electricity. Only black holes have more potential as a reservoir for free energy, releasing around 100\,MeV per nucleon that is accreted, but that energy remains largely untapped.

\subsection{Coda}

I sang praises of big-bang nucleosynthesis in a {\it Physics Today} article I wrote in 2008 which celebrated the 60th anniversary of the $\alpha\beta\gamma$ paper.  Here, I have tried to elucidate the basic physics underpinning BBN and clarify the history of that physics.  

Further, BBN serves as a case study of how big ideas develop.  This big idea began with the lofty goal of explaining the origin of the elements in the periodic table.  In the end, only the lightest elements were produced in the big bang.  However, the paper opened the door to the hot, early phase of the Universe where particle interactions shaped the Universe that we see today, and it played a role in inventing the modern r-process theory of the heavy elements.  While $\alpha\beta\gamma$ did not achieve its original goal, it spurred the thinking that achieved even more transformational advances in cosmology.

I mention two other important discussions of the history of BBN physics, by Peebles in his most recent book \cite{PJEPCC}, {\it Cosmology's Century}, and by Alpher and Herman in their 1988 {\it Physics Today} article \cite{AlpherHermanPT}.  
In his book, Peebles devotes more than 20 pages to discussing much of the same physics history and many of the same papers as I do here (pp. 130 to 158).  

Peebles' account has a strong personal and astronomical bent -  he was of course a central participant - with interesting reminiscences and personal correspondence and commentary on the astronomical observations.  In contrast, I have tried to make my through line the physics and how it evolved.  I hope my paper is a valuable complement to Peebles' account.

In their {\it Physics Today} article, Alpher and Herman reflect upon their BBN work and reminisce about their collaboration with a larger than life figure, George Gamow.  Their look back, and another one by Alpher's son \cite{VAlpher}, describe the joys and difficulties of working with and in the shadow of such a prominent and creative physicist.  The key idea behind BBN was Gamow's \cite{Gamow46}, and his involvement brought great attention to their work.  At the same time, he captured much of the credit, e.g., it was in his sole author {\it Nature} paper \cite{gamow48} where the results of Alpher's thesis were first shown.  It may have also been Gamow's influence that kept Alpher and Herman so closely wedded to the non-equilibrium model.  And what comes out most clearly is their disappointment at not getting more credit for their contributions.  It took a village to create the modern theory of BBN and they were an important part of the village.

\vskip 20 pt

I thank Robert Wagoner and Robert Scherrer for their comments and insightful conversations and Helge Kragh for his constructive criticisms.

\begin{appendices}

\section{Some technical details}\label{secA1}
Here I add a few more details underpinning my estimates in  \S2, with the key formulae and approximations used in my calculations.  These are the formulae for the expansion rate $H$, the age of the Universe $t$, the energy density in relativistic particles, reaction rates, and the number density of photons and baryons:
\begin{eqnarray}
H^2 &\equiv& \left({\dot a \over a}\right)^2 = {8\pi G \rho \over 3} \simeq{8\pi G \rho_R \over 3}  \nonumber \\
\rho_R & =  & g_*{\pi^2 \over 30} T^4  \nonumber \\
H & \simeq & 10 T^2/m_{\rm pl}  \nonumber \\
t & = & 1/2H \simeq { 1\,{\rm sec} \over (T/{\rm MeV})^2} \simeq  { 1\,{\rm sec} \over (T/10^{10}{\rm \, K})^2}\nonumber  \\
\Gamma & = & n <\sigma v >  \nonumber 
%%n_\gamma & = & {2 \zeta (3) \over \pi^2} T^3 \nonumber  \\
%%n_B & \simeq & \eta n_\gamma  
\end{eqnarray}
where reaction rate (per particle per second) $\Gamma$ is proportional to the number density of targets times the thermally-averaged cross section times relative velocity.
Throughout $k_B = \hbar = c =1$ so that $G=1/m_{\rm pl}^2$ ($m_{\rm pl} = 1.22 \times 10^{19}\,$GeV is the Planck mass); $g_*(T)\simeq 3-300$ counts the number of relativistic degrees of freedom at temperature $T$.  
During BBN the energy density is dominated by that in the thermal bath of relativistic particles (photons, neutrinos, antineutrinos and electron-positron pairs) and $g_* = 10.75$, so that $H \simeq 10 T^2/m_{\rm pl}$.

Crucial to understanding BBN is Nuclear Statistical Equilibrium (NSE).  It means that the various nuclides are in full thermal equilibrium, with number densities in the non-relativistic limit given by,
\begin{eqnarray}
 n_A & = & \left( {m_AT \over 2\pi }\right)^{3/2} \exp \left( {\mu_A - m_A \over T } \right) , \\
 B_A & \equiv & Z m_p + (A-Z)m_n - m_A, 
\end{eqnarray}
where $m_A$ and $\mu_A$ are the mass and chemical potential of nuclide $A$, $m_p$, $m_n$ are the proton and neutron masses respectively, and $\mu_A = Z\mu_p + (A-Z) \mu_n$.  NSE  determines when nuclei are favored over free nucleons, through the competition between the lower energy of nuclei specified by $B_A$ and the higher entropy of $A$ free nucleons, cf. Eq.~\ref{T_nuclei} in \S2:
\begin{eqnarray}
    T_{\rm nuclei} & \simeq & { B_A/(A-1) \over \ln \eta^{-1} + 1.5 \ln (m/T_{\rm nuclei}) }\, . \nonumber
    \label{T_nuclei}
\end{eqnarray}

A key concept of early Universe cosmology is the ``freeze out" of particle interactions.  Namely, when a reaction rate key to adjusting a particle abundance cannot keep up with the rate at which the temperature is falling, $\dot T/T = -\dot a /a = H$, it ceases to be relevant and the abundance it is controlling ``freezes in.''  To a good approximation, a reaction rate is frozen out when $\Gamma < H$, and freeze out occurs when $\Gamma \simeq H$.  

For example, the rate for the reactions that maintain thermal balance of neutrons and protons, $p + e^- \longleftrightarrow n + \nu_e $ etc, is $\Gamma \sim G_F^2T^5$.  Comparing this rate to the expansion rate, it follows that the freeze of the neutron-to-proton ratio occurs 
\begin{eqnarray}
T_F \simeq G_F^{-2/3} m_{\rm pl}^{-1/3} \sim 1\,{\rm MeV}\,.  
\end{eqnarray}

The rate of any strong or electromagnetic (EM) charged-particle nuclear interaction will have a Coulomb barrier suppression factor of the form,
\begin{eqnarray}
    \exp [-2{\bar A}^{1/3} (Z_1 Z_2)^{2/3} /(T/{\rm MeV})^{1/3} ]\, ,
\end{eqnarray} 
with an overall nuclear cross section size, $m_\pi ^{-2} \simeq 2\times 10^{-26}\,{\rm cm}^2$, and an additional factor of $\alpha_{EM}/\pi$ for an EM interaction.  The controlling factor for freeze out is the exponential Coulomb barrier factor. And the Coulomb barrier is lowest for charge +1 on charge +1 and small $\bar A \equiv A_1A_2/(A_1+A_2)$.  ${\bar A}$ is smallest for $A_1 = 1$ and $A_2 = 2$ ($=2/3$) and $A_1 = A_2 =2$ ($= 1$).  To estimate a charged-particle nuclear rate, I have used $d + d \rightarrow {^4{\rm He}} + \gamma$,
\begin{eqnarray}
   \Gamma \sim  \eta T^3 {\alpha_{ EM} /\pi \over {m_\pi}^2 } \exp [ -2(kT/{\rm MeV})^{-1/3} ]\ .
\end{eqnarray}
Comparing this rate to the expansion rate leads to Eq.~(2).

For a more detailed and expansive discussion of BBN and particle interactions in the early Universe see Refs.~\cite{KolbTurner,Weinberg72}.

\section{The Alpher-Herman 5\,K cosmology}\label{secAH}

The Alpher-Herman prediction of a CMB temperature of about 5\,K \cite{alpherherman48} is predicated upon their negatively-curved cosmological model, which was inspired by Gamow's update to $\alpha\beta\gamma$ \cite{gamow48}.  In Gamow's update, he adds numerical results from Alpher's Ph.D. thesis \cite{alpher48} and a new twist, an estimate of the Jeans mass at matter/radiation equality, $3\times 10^7\,M_{\odot}$, connecting its value to structure formation.

Alpher and Herman \cite{alpherherman48} add negative curvature to avoid having a radiation-dominated Universe today.  For the parameters they infer from the neutron-capture model for BBN \cite{gamow48,alpherherman48}, they find matter-radiation equality occurs at $t = 30\,$Gyr or about ten times the present age of the Universe (as estimated in 1948) and a temperature of $16\,$K.  

Adding negative curvature can avoid that fate, and the equation for the evolution of the scale factor $a(t)$ becomes
\begin{eqnarray}
   H^2 \equiv \left( {\dot a \over a} \right)^2 = {8\pi G  \left( \rho_\gamma + \rho_B \right) \over 3} - {k \over a^2} \, ,
\label{FRW}   
\end{eqnarray}
where $k$ is the curvature (negative for negatively-curved models), $\rho_\gamma$ is the photon mass density and $\rho_B$ is the matter density; cf., their equation at the bottom of page 774 \cite{alpherherman48}.

\begin{figure}
\centering
  \includegraphics[width=0.9\linewidth]{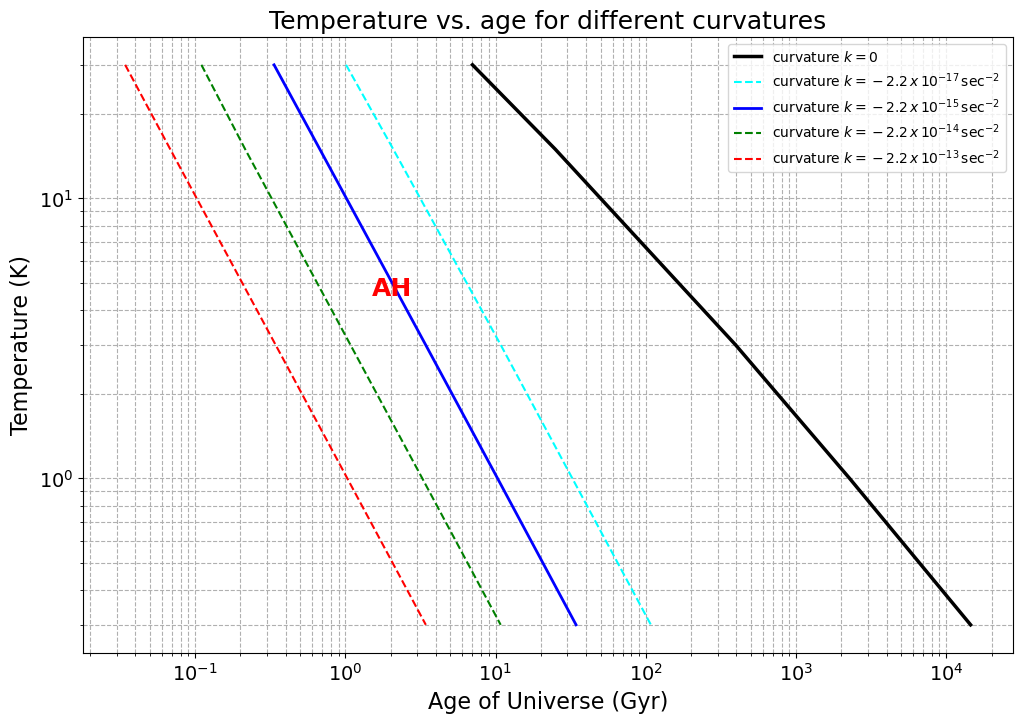}
  \caption{The relationship between the temperature and the age of the Universe for different values of the curvature $k$ in the Alpher-Herman scenario.  The symbol ``AH'' indicates the model they chose, which leads to their prediction of 5\,K.  Note, since the high-curvature models shown evolve as $a(t) \propto t$, the temperature and age are related by $T \propto t$.  For the zero-curvature model and $T> 16\,$K or so, the Universe is radiation-dominated and $a(t) \propto t^{1/2}$, so that the age and temperature are related by $T \propto t^{-1/2}$; for lower temperatures it is matter-dominated and $a(t) \propto t^{2/3}$, so that the age and temperature are related by $T \propto t^{-2/3}$.} \label{fig:AlpherHerman}
\end{figure}

Based upon their neutron-capture model calculations they fix the photon temperature to be $1.52\times 10^{10}\,$K at 1 second and the matter density to be $4.83 \times 10^{-4}\,{\rm g\,cm}^{-3}$ at the same time.  This leads to a one-parameter family of models - the curvature $k$ - for the relationship between the age of the Universe and its temperature.  Fig.~\ref{fig:AlpherHerman} illustrates this family of models, based upon the results of integrating Eq.~\ref{FRW}.

Their prediction of the CMB temperature today proceeds differently from that using the BBN-preferred value (or range) of $\eta$ and the present matter density to predict the present temperature discussed in \S\ref{counterfactual}.  That procedure does not depend upon the cosmological model; only the constancy of $\eta$ from BBN until today.

According to the text at the top of page 775 \cite{alpherherman48}, here is what they did:  they chose their model ($k=-2.2\times 10^{-15}\,{\rm sec}^{-2}$)\footnote{They use $\ell$ to denote the scale factor and specify the curvature by $R_0 = ca_0|k|^{-1/2}$.} by requiring that the Jeans mass at matter/radiation equality correspond to the value that Gamow obtained in Ref.~\cite{gamow48}, $3\times 10^7\,M_\odot$, and then defined the present epoch by its age of 2\,Gyr.  This then leads to their prediction of $T= 5\,$K.  

There is only one value of the curvature that results in a 2\,Gyr old universe having a temperature of 5\,K ($k=-2.2\times 10^{-15}\,{\rm sec}^{-2}$), and so I am confident that I have the model they were discussing.  That being said, I obtain different numbers for the age of the Universe at matter/radiation equality, $t_{EQ} = 6.3\times 10^{16}\,$sec vs. their $3.5\times 10^{14}\,$sec, and more importantly a different temperature, $T_{EQ} = 16\,$K vs. their 600\,K.  The latter is more important because, $T_{EQ}$ does not depend upon the cosmological model, only upon the temperature and matter density specified at 1\,second; 16\,K is the correct value for matter/radiation equality, calling into question their results for $t_{EQ}$.

A few comments about their 5\,K cosmology.  The matter density $\rho_B = 1.8 \times 10^{-32}\,{\rm g\,cm}^{-3}$ is very low ($\Omega_B = 4\times 10^{-5}$), and the dynamics of the expansion have been curvature-dominated since $t = 9.2\times 10^{13}\,$sec,  with $a(t) \propto t$.  While the Hubble constant is very large, $H_0 = 488\,$km/s/Mpc and the age corresponding low, $t = 2\,$Gyr, both were consistent with estimates at the time.

An aside, had they not added curvature and used the age of the Universe - about 2\,Gyr - to fix the present temperature and matter density, they would have predicted a CMB temperature of around 60\,K and a matter density of around $3\times 10^{-29}\,{\rm g\, cm}^{-3}$.

To sum up, their prediction of $T_{\rm CMB} = 5\,$K is certainly fortuitous since it is based upon the wrong nuclear physics, an odd argument argument about the Jeans mass, a numerical error, and a strange cosmological model.




\end{appendices}

%%\newpage

\section*{Declarations}

\begin{itemize}
\item Funding:  Not applicable
\item Availability of data and material:  Not applicable
\item Competing interests: Not applicable
\item Author contributions: Not applicable
\end{itemize}

%%\bibliography{sn-bibliography}% common bib file

\end{document}